\documentclass[aps,prl,superscriptaddress,floatfix,reprint,showkeys,showpacs]{revtex4-1}

\usepackage{graphicx}
\usepackage{subfigure}
\usepackage{float}
\usepackage{bm}
\usepackage[space]{grffile}
\usepackage{leftidx}
\usepackage{amsmath,amssymb,amstext}
\usepackage{xcolor}
\usepackage{wrapfig}
\usepackage{physics}
\usepackage{hhline}
\usepackage{hyperref}
\usepackage{braket}
\usepackage{amsthm}
\usepackage{footnote}

\def\Rb{$^{87}\textrm{Rb}\,$}

\newcommand{\gse}{\Gamma_\mathrm{se}}
\newcommand{\gsm}{\Gamma_\mathrm{sm}}
\newcommand{\gqe}{\Gamma_\mathrm{qe}}
\def\m1cond{$m_F{=}-1$}

\begin{document}
\title{Collisional spin transfer in an atomic heteronuclear spinor Bose gas}
\author{Fang Fang}
\email{akiraff@berkeley.edu}
\affiliation{Department of Physics, University of California, Berkeley, California 94720, USA}
\author{Joshua A. Isaacs}
\affiliation{Department of Physics, University of California, Berkeley, California 94720, USA}
\author{Aaron Smull}
\affiliation{Department of Physics, University of California, Berkeley, California 94720, USA}
\author{Katinka Horn}
\email{Current address: Department of Physics, Universität Hamburg, 22761  
Hamburg, Germany}
\affiliation{Department of Physics, University of California, Berkeley, California 94720, USA}
\author{L.Dalila Robledo-De Basabe}
\affiliation{Department of Physics, University of California, Berkeley, California 94720, USA}
\author{Yimeng Wang}
\affiliation{Department of Physics and Astronomy, Purdue University, West Lafayette, Indiana 47907, USA }
\author{Chris H. Greene}
\affiliation{Department of Physics and Astronomy, Purdue University, West Lafayette, Indiana 47907, USA }
\affiliation{Purdue Quantum Science and Engineering Institute, Purdue University, West Lafayette, Indiana 47907, USA }
\author{Dan M. Stamper-Kurn}
\affiliation{Department of Physics, University of California, Berkeley, California 94720, USA}
\affiliation{Materials Sciences Division, Lawrence Berkeley National Laboratory, Berkeley, California 94720, USA}

\begin{abstract}
We observe spin transfer within a non-degenerate heteronuclear spinor atomic gas comprised of a small $^7$Li population admixed with a $^{87}$Rb bath, with both elements in their $F=1$ hyperfine spin manifolds.  Prepared in a non-equilibrium initial state, the $^7$Li spin distribution evolves through incoherent spin-changing collisions toward a steady-state distribution.  We identify and measure the cross-sections of all three types of spin-dependent heteronuclear collisions, namely the spin-exchange, spin-mixing, and quadrupole-exchange interactions, and find agreement with predictions of heteronuclear $^7$Li-$^{87}$Rb interactions at low energy.  Moreover, we observe that the steady state of the $^7$Li spinor gas can be controlled by varying the composition of the $^{87}$Rb spin bath with which it interacts.

\end{abstract}
\maketitle

The spin dynamics of a gas of itinerant particles are central to a  range of phenomena in atomic and condensed-matter physics.  In solid state, spin relaxation governs the penetration depth of non-equilibrium electron spin distributions injected from ferromagnetic into non-magnetic materials \cite{aron76injection,john85interfacial}, affecting spintronics and their applications in magnetic data storage and sensing.  The spin-dependent exchange interaction generates spin waves in atomic Bose  \cite{bige89, mcgu02} and Fermi gases \cite{du09} and in electronic systems \cite{aron77spinwave}.  Spin relaxation collisions also regulate demagnetization cooling of atomic gases \cite{fatt06demag,medl11demag,nayl15demag,olf15cooling}.


Optical trapping of atomic gases enables the study of spinor Bose and Fermi gases, in which atoms may populate any of the magnetic sublevels within a manifold of fixed total atomic spin \cite{stam13rmp}.  These gases present a controlled medium for studying magnetic ordering, phase transitions, and non-equilibrium spin dynamics.
Recently, studies have extended to heteronuclear spinor gases, comprised of spinor gases of two different elements or isotopes \cite{li15heter}.

\begin{figure}[t]
	\begin{center}
		\includegraphics[width=0.48\textwidth]{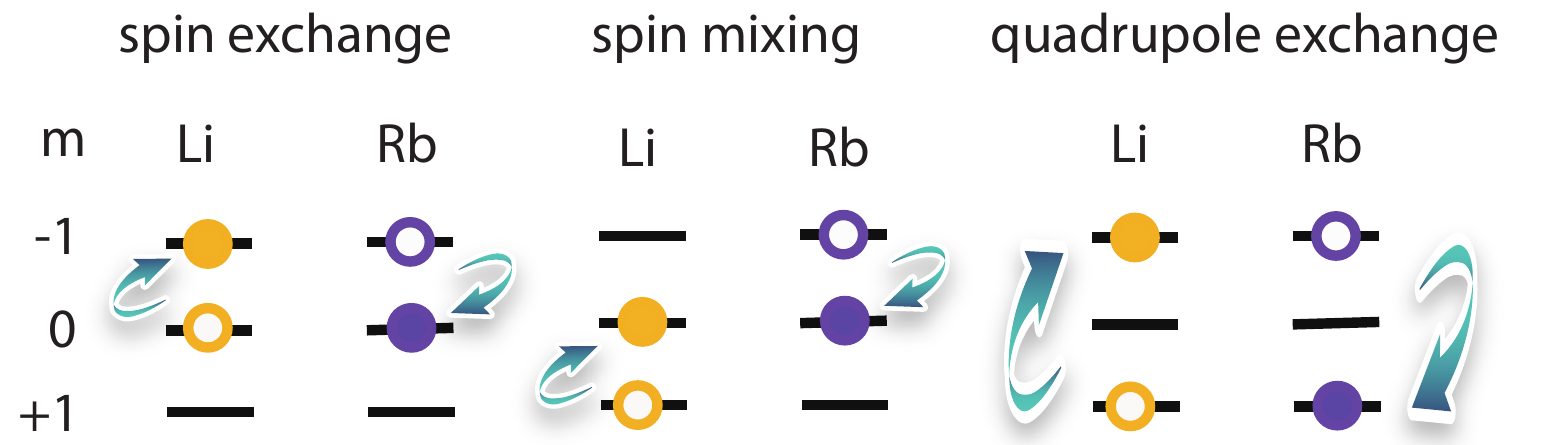}
	     \caption{A heteronuclear spinor gas, with both spinor gases in their respective $F=1$ spin manifolds, undergoes three distinct  spin-changing s-wave collisions: spin exchange, spin mixing, and quadrupole exchange.  One possible channel is shown for each collision, with initial and final Zeeman states indicated by an open and a closed symbol, respectively.  We assume spinor-gas collisions respect rotational symmetry and ignore effects of magnetic dipole-dipole interactions. }
\label{fig:SpinschemeNew.pdf}
\end{center}
 \end{figure}

Interactions in spinor gases are governed by rotational symmetry. Further, neglecting effects of magnetic dipole-dipole interactions, heteronuclear s-wave collisions  preserve the total spin projection of the colliding atom pair along a chosen quantization axis (defined as $\mathbf{z}$).  In a homonuclear $F=1$ spinor gas, these constraints, together with Bose-Einstein statistics,  leave spin-mixing as the only channel for spin-changing collisions.  The dynamics of a heteronuclear spinor gas are richer. 
Considering a mixture of $F=1$ spinor gases of elements $A$ and $B$, and labeling the state of an $A$-$B$ atom pair by the magnetic quantum number $m_z$ of each atom, we identify three distinct processes that transfer spin between the two spinor gases (Fig.\ \ref{fig:SpinschemeNew.pdf}): spin exchange (exemplified by the process $|0_A,-1_B\rangle \leftrightarrow |-1_A, 0_B\rangle$) and spin mixing (e.g.\  $|-1_A,+1_B\rangle \leftrightarrow |0_A,0_B\rangle$), both of which transfer one quantum of spin between the spinor gases; and quadrupole exchange (e.g.\  $|+1_A,-1_B\rangle \leftrightarrow |-1_A,+1_B\rangle$), which transfers two spin quanta.
Experiments to date have focused only on spin exchange within $^{23}$Na-$^{87}$Rb \cite{li15heter} and $^7$Li-$^{23}$Na \cite{Mil1128} heteronuclear spinor gases.


In this work, we characterize spin relaxation in the heteronuclear $^7$Li-$^{87}$Rb spinor gas.  We prepare non-degenerate  gas mixtures with a large number  imbalance, so that the majority population of Rb atoms serves as a spin bath of near-constant Zeeman-state distribution, and then observe how the minority population of Li atoms evolves through incoherent spin-changing collisions with this bath.  To make an analogy with spintronics, Rb plays the role of a magnetic material and Li plays the role of an injected non-equilibrium electron current.   By comparing spin relaxation from several initial states, we identify and characterize the strength of all three spin-changing processes, obtaining the first complete characterization of spin-dependent interactions in a heteronuclear spinor gas. 

Li and Rb atoms are prepared in a two-species magneto-optical trap (MOT) \cite{fang20cross}. During the final 0.5 s of loading this MOT, an optical dipole trap is introduced, formed by a single light beam of 1064-nm wavelength and focused to a $1/e^2$ beam radius of 34 $\mu$m at the MOT center.  This optical trap has trap depth of $k_B \times 330 \, \mu$K and frequencies $\omega_{x,y,z} = 2 \pi \times (5700, 5700, 34) \, \mbox{s}^{-1}$ for Li, and of $k_B \times 840\, \mu$K and $\omega_{x,y,z} = 2 \pi \times (2600, 2600, 16) \, \mbox{s}^{-1}$ for Rb.  After the MOT is turned off, a final stage of optical cooling leaves an optically trapped gas of about 2 $\times$ $10^5$ $^7$Li atoms at a temperature of $T_\mathrm{Li} = 36\, \mu$K, and about 2 $\times$ $10^6$ $^{87}$Rb atoms at a temperature of $T_\mathrm{Rb} = 87\, \mu$K.  Since the ratio of the temperature to the optical trap depth are similar for the two elements, they occupy the same trap volume. Differential effects of gravity (oriented along $\mathbf{y}$) are negligible.

\begin{figure*}[t]
	\begin{center}
		\includegraphics[width=0.95\textwidth]{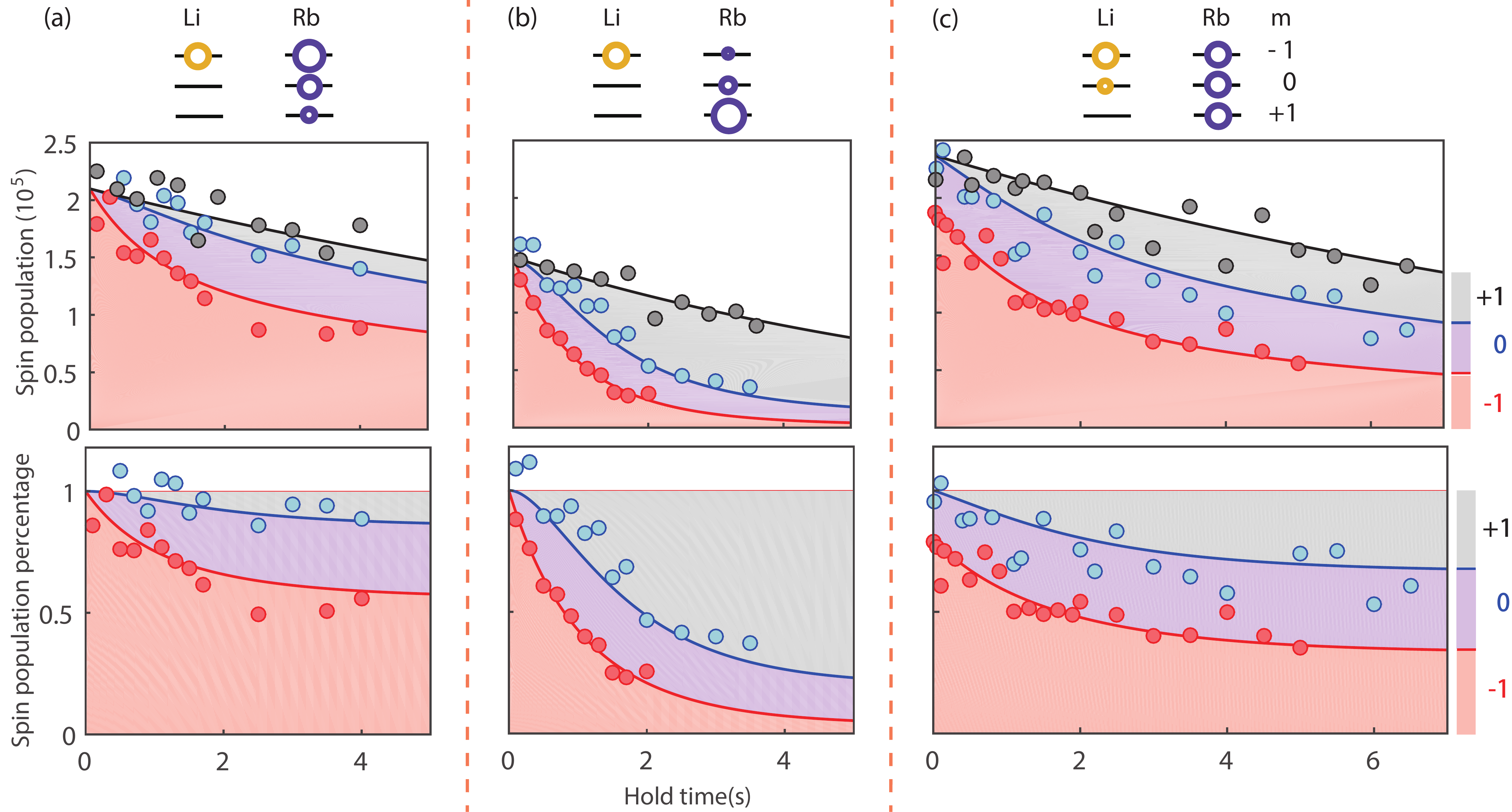}
	     \caption{Spin relaxation of a $^7$Li gas in the presence of a $^{87}$Rb spin bath.  Top row: The Li and Rb gases are prepared in one of three initial spin configurations (a-c), with Zeeman-state distributions indicated graphically.  Measurements of $N^{(\mathrm{Li})}_{-1}$ (red), $N^{(\mathrm{Li})}_{0,-1}$ (blue) and $N^{(\mathrm{Li})}_\mathrm{tot}$ (gray) and the corresponding fractional populations are shown in the middle and bottom row, respectively. Here, $N^{(\mathrm{Li})}_{0,-1}$ is the sum of the atom population in $|m_F = -1\rangle$ and $|m_F = 0\rangle$ states. Data are overlaid with results of a fit of all measurements to a rate equation (Eq.\ \ref{eq:rate3}).  Shaded areas indicate Li populations in the $m_F = -1$ (red), $0$ (purple), and $+1$ (gray) sublevels.  The Rb atom number is $1.6 \times 10^6$ in (a) and (c), and $2 \times 10^6$ in (b).}
\label{fig:LiRelaxInRbBath.pdf}
\end{center}
 \end{figure*}

The two spinor gases are initially unpolarized, with equal population in all three magnetic sublevels of the $F=1$ spin manifolds and no transverse coherence between sublevels. To initiate spin dynamics, we polarize the gas through optical pumping on either the D1 (Li) or D2 (Rb) principal atomic resonances.  Pumping with $\sigma_\pm$ circular polarized light propagating in the direction of an applied magnetic field, we drive either element selectively into the $|m_F = \pm 1\rangle$  state with  fidelity of nearly 100\% for Li and around 80\% for Rb.  A wider range of initial states can be accessed by applying rf magnetic field pulses resonant with the Larmor frequency.  For example, by optically pumping the Rb gas, applying a resonant rf pulse with pulse angle $\phi$, and then optically pumping the Li gas, we prepare the initial spin configuration shown in Fig.\ \ref{fig:LiRelaxInRbBath.pdf}(a), with Li occupying the $|m_F = -1\rangle$ state and the Rb bath occupying a variable mixture of magnetic sublevels.
 
The atoms are allowed to evolve for up to several seconds before the spin composition of each gas is measured.  We measure the total $N^{(\mathrm{Rb})}_{m_F}$ and fractional $p^{(\mathrm{Rb})}_{m_F}$ Rb populations in each magnetic sublevel using state-selective absorption imaging \cite{rama12partial,mart14magnon,suppmat}.

We quantify the Li spin distribution differently. Prior to imaging, we impose a spherical quadrupole magnetic field (with axis along $\mathbf{y}$) atop the optical trap \footnote{$B^\prime_y$ is ramped up at about 10 G/cm per ms, held at the target value for about 40 ms, and then ramped off before imaging}.  At a low magnetic field gradient ($B^\prime_y = 58$ G/cm), magnetic forces expel all lithium atoms in the $|m_F = +1\rangle$ sublevel, but allow atoms in the remaining sublevels to remain trapped.  At a higher gradient ($B^\prime_y = 270$ G/cm) the quadratic Zeeman shift causes also the $m_F = 0$ atoms to be expelled, leaving only the weak-field seeking $m_F = -1$ atoms. By counting the remaining trapped Li gas through spin-independent absorption imaging, we determine the total atom number either in all sublevels $N^{(\mathrm{Li})}_\mathrm{tot}$, in both the $|m_F = 0\rangle$ and $|m_F = -1\rangle$ states $N^{(\mathrm{Li})}_{0,-1}$, or in just the $|m_F = -1\rangle$ state $N^{(\mathrm{Li})}_{-1}$, depending on whether we had previously applied no gradient, a low gradient, or a high gradient, respectively.

In each repetition of the experiment, we make just one such measurements on just one of the two co-trapped elements.  The full spin distribution is obtained by combining measurements from several repetitions of the same experimental sequence.

As shown in Fig.\ \ref{fig:LiRelaxInRbBath.pdf}, an initially spin polarized Li gas undergoes spin relaxation over the course of several seconds, and both the nature and final state of this relaxation varies as the spin polarization of the Rb spin bath is varied.  We present several arguments for why the relaxation occurs solely due to spin-dependent Li-Rb collisions.  First, we conducted separate experiments in which a single-species gas of either Li or Rb was found to remain spin polarized in the optical trap, demonstrating that spin relaxation is not caused by external factors such as magnetic field noise or light scattering.  Second, we may discount the effect of spin-changing Li-Li collisions, relative to that of Li-Rb collisions, given that $N^{(\mathrm{Rb})}_\mathrm{tot}  \gg N^{(\mathrm{Li})}_\mathrm{tot}$.  Third, we may neglect spin-changing Rb-Rb collisions over the measurement duration because the rate of spin-mixing collisions in $^{87}$Rb $F=1$ spinor gases is anomalously small \cite{klau01rbspin}.

We compare the observed spin relaxation to a model for spin-dependent interactions in a heteronuclear spinor gas.  We distinguish between the spin-dependent effects of forward scattering, in which the momenta of the colliding particles do not change in a collision, and non-forward scattering, in which they do.  Forward scattering leads to collective mean-field  effects such as Zeeman-state-dependent energy shifts and to dynamics where spin is coherently transferred between the two spinor gases.  These coherent dynamics, observed previously in quantum-degenerate or near quantum-degenerate heteronuclear spinor gases \cite{li15heter,Mil1128}, are driven by transverse coherence between Zeeman populations.  In our system, however, such transverse coherence is damped rapidly, within just a few ms, owing to magnetic field inhomogeneity and atomic motion   \cite{higb05larmor,ramseynote}, and, thus, coherent spin transfer is suppressed.

Non-forward scattering, in a non-degenerate gas,  leads only to incoherent transfer of spin between spinor gases.  Considering a uniform mixture of spinor gases $A$ and $B$, the per-atom rate at which atoms in gas $A$ and sublevel $m_A$ undergo a spin-changing collision is given as $\Gamma_\alpha p^{(B)}_{m_B} = n^{(B)} \bar{v} \sigma_\alpha p^{(B)}_{m_B}$ where $n^{(B)}$ is the density of gas $B$ atoms, $\bar{v}$ is the rms incident velocity,  $\sigma_\alpha$ is the cross section, and $p^{(B)}_{m_B}$ is the fraction of gas $B$ atoms in the incident spin state for the specific spin-changing collision considered (labeled by $\alpha$) \footnote{This expression for $\Gamma_\alpha$ assumes the total internal-state energy to be the same before and after the collision.  In our experiment, small energy differences do arise from the small gyromagnetic ratio difference between $^7$Li and $^{87}$Rb and from mean-field energy shifts. However, these energy differences ($|\Delta E| \lesssim k_B \times 1$ nK) are far smaller than $k_B T_\mathrm{Li}$ and $k_B T_\mathrm{Rb}$, and can thus be ignored.}.

Following these considerations, we describe the evolution of spin populations in our Li gas, $\mathbf{N}^{(\mathrm{Li})} = \left(N^{(\mathrm{Li})}_{+1}, N^{(\mathrm{Li})}_{0}, N^{(\mathrm{Li})}_{-1}\right)^\intercal$ through the rate equation
\begin{widetext}
\begin{equation}
        \frac{d \mathbf{N}^{(\mathrm{Li})}}{d t} = -\frac{\mathbf{N}^{(\mathrm{Li})}}{\tau} + \left(\begin{array}{c c c} -\gse p_0 - \left(\gsm + \gqe\right) p_{-1} & \gse p_{+1} + \gsm p_0 & \gqe p_{+1} \\
        \gse p_0 + \gsm p_{-1} & -\gse \left(p_{+1} + p_{-1}\right) - 2 \gsm p_0 & \gse p_0 + \gsm p_{+1} \\
        \gqe p_{-1} & \gse p_{-1} + \gsm p_0 & -\gse p_0 - \left(\gsm + \gqe\right) p_{+1} \end{array}   \right) \mathbf{N}^{(\mathrm{Li})}
    \label{eq:rate3}
\end{equation}
\end{widetext}
Here, $\gse$, $\gsm$ and $\gqe$ quantify the rates of spin-exchange, spin-mixing, and quadrupole-exchange collisions, respectively.  We make the simplifying assumption that the total atom number and Zeeman-state populations $p^{(\mathrm{Rb})}_{m}=p_m$ of the Rb bath remain constant.
We include also a spin-independent loss term reflecting the preferential loss of Li atoms from the optical trap, with trap lifetime $\tau$, over the experimental duration.

This rate equation ignores possible variations of the Li and Rb spin distributions in position, momentum and energy.  We may neglect these variations owing to rapid atomic motion as compared to the slow Li-Rb collision rates, and to the fact that non-forward scattering leads to effective redistribution of the Li-atom momentum.

Using Eq.\ \ref{eq:rate3}, we apply a simultaneous least-squares fit to all data in Fig.\ \ref{fig:LiRelaxInRbBath.pdf}.  The cross-sections $\sigma_\mathrm{se}$, $\sigma_\mathrm{sm}$ and $\sigma_\mathrm{qe}$, and also the initial total Li atom number at each experimental setting, are treated as free parameters.  The uniform Rb density in our expression for $\Gamma_\alpha$ is replaced by the overlap density (around $1.5 \times 10^{12} \, \mbox{cm}^{-3}$) determined separately for each initial condition \footnote{We find the gas temperatures to remain fairly constant and independent of spin state. The lack of thermalization between the Li and Rb gases is expected given the low collision rates and the large difference in atomic masses.}.  The relative velocity is calculated as a thermal average: $\bar{v} = \sqrt{\frac{8k_B}{\pi}\left(\frac{T_\mathrm{Rb}}{m_\mathrm{Rb}}+\frac{T_\mathrm{Li}}{m_\mathrm{Li}}\right)} \simeq 0.36 \, \mbox{m}/\mbox{s}$, where  $m_\mathrm{Rb}$ and $m_\mathrm{Li}$ are atomic masses.

Different initial conditions provide varying constraints on the measured cross-sections. For example, for initial conditions (a) (see Fig.\ \ref{fig:LiRelaxInRbBath.pdf}), the observed spin relaxation of Li from the initial spin-polarized $|m_F = -1\rangle$ state is dominated by the spin-exchange interaction, i.e.\ through collisions with the Rb $m_F = 0$ population.  Alternately, for initial condition (b), the $|m_F = -1\rangle$ state of Li initially undergoes either spin-mixing or quadrupole-exchange interactions with the Rb bath that is mostly in the $|m_F = +1\rangle$ state.  The fact that we observe an immediate, linear growth of the Li population in the $|m_F = 0\rangle$ state, followed more slowly by a rise in the $|m_F = 1\rangle$ population, shows the spin-mixing process to dominate over the quadrupole-exchange process, i.e.\  $\sigma_\mathrm{sm}\gg\sigma_\mathrm{qe}$. The fact that spin relaxation from initial conditions (a) and (b) occurs at a comparable rate demonstrates that $\sigma_\mathrm{se}$ and $\sigma_\mathrm{sm}$ are comparable in value.  The spin-dependent collision cross-sections determined from our fits, and shown in Table \ref{Table:Scatter}, reflect these observations.

\begin{table}[t]
\begin{tabular}{|c|c|c|}
  \hline
  Quantity  & Experiment & Theory \\
  \hhline{|-|-|-|}
 $\sigma_\mathrm{se}$ ($10^{-14}\, \mathrm{cm^2}$) & $2.4(7)$ & $2.2$ \\
  \hhline{|-|-|-|}
 $\sigma_\mathrm{sm}$ ($10^{-14}\, \mathrm{cm^2}$) & $1.7(6)$ & $1.8$ \\
  \hhline{|-|-|-|}
$\sigma_\mathrm{qe}$ $^\mathrm{(ub)}_\mathrm{(lb)}$ ($10^{-14} \, \mathrm{cm^2}$) & $0$ $^{(0.5)}_{(0)}$ & 0.02 \\
  \hhline{|-|-|-|}
$\sigma_\mathrm{se}/\sigma_\mathrm{sm}$ & $1.4(6)$ & $1.2$  \\ 
\hhline{|-|-|-|}
 $| a_\mathrm{se} | (a_B)$ & $ 8.2(1.2) $ & $7.8$ \\
\hhline{|-|-|-|}
 $| a_\mathrm{sm} | (a_B)$ & $ 7.0(1.3) $ & $7.2$ \\
\hhline{|-|-|-|}
$| a_\mathrm{qe}|$ $^\mathrm{(ub)}_\mathrm{(lb)}$  $(a_B)$ & $ 0$ $^{(3.8)}_{(0)}$ & 0.7 \\
  \hline
\end{tabular}
\caption{Heteronuclear spin-dependent interaction strengths for the $^7$Li-$^{87}$Rb spinor gas.  Cross sections and effective scattering lengths obtained from experimental fits are compared to theoretical predictions.  For experimental data, standard statistical errors, accounting for fitting errors and trap frequency, temperature, and Rb atom number measurements, are shown for the spin-exchange and spin-mixing processes, while standard upper (ub) and lower (lb) bounds are shown for the quandrupole-exchange process.}
\label{Table:Scatter}
\end{table}

Low energy collisions between two spinor Bose gases are characterized fully by the s-wave scattering lengths $a_{F_\mathrm{tot}}$ labeled by the total angular momentum quantum number $F_\mathrm{tot}$ for the colliding atoms; here, $F_\mathrm{tot} \in \{ 0,1, 2\}$. Effective scattering lengths $a_\mathrm{eff}$ for the three allowed spin-changing collisions are obtained as differences between the values of $a_{F_\mathrm{tot}}$, with $a_\mathrm{se} = (a_2-a_1)/2$, $a_\mathrm{sm} = (a_2 - a_0)/3$ and $a_\mathrm{qe} = (a_2 - 3a_1 + 2a_0)/6$ for the spin-exchange, spin-mixing, and quadrupole-exchange processes, respectively.  Collision cross sections for each process are calculated  as $\sigma= 4 \pi a_\mathrm{eff}^2$ \cite{suppmat}.

We compute the scattering lengths $a_{F_\mathrm{tot}}$ through a coupled-channel scattering calculation that relies on the model of the $^7$Li-$^{87}$Rb molecular potential in Ref.\ \cite{maie15efimov}.  Calculations assume a collision energy of $k_B \times \, 50 \, \mu$K.  As shown in Table \ref{Table:Scatter}, our measurements of the spin-changing collision cross sections and effective scattering lengths are consistent with these predictions. In spite of this agreement with the predicted differences in the scattering lengths $a_{F_\mathrm{tot}}$, we note that previous measurements found that the mean value of these lengths differs from theoretical predictions at collision energies of 100's of $\mu$K \cite{fang20cross}.  We note also that our calculations predict the effective scattering lengths to be relatively independent of collision energy within the range probed in our present experiment.

Following several seconds of spin relaxation, the Li gas evolves to a spin-state distribution that depends on the state of the Rb bath.  This situation bears analogy to spintronic systems, where an electron current acquires magnetization through the spin-transfer torque effect after being injected into a ferromagnetic material. 
This steady-state Li spin distribution is determined by the non-trivial zero-eigenvalue eigenvector of the matrix in Eq.\ \ref{eq:rate3}.

\begin{figure}[t]
	\begin{center}
		\includegraphics[width=0.40\textwidth]{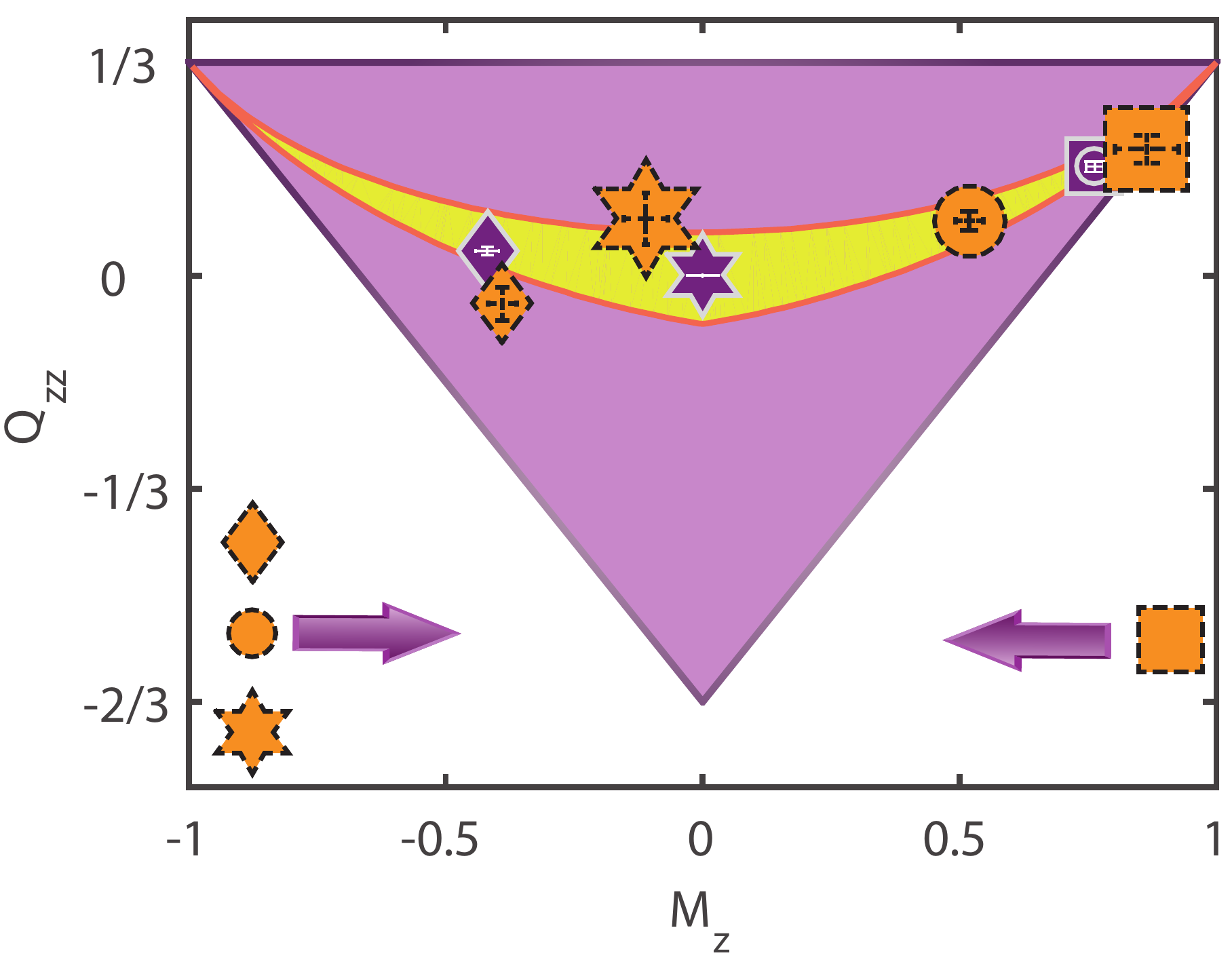}
	     \caption{Controlling the steady-state spin distribution of the $^7$Li spinor gas by varying the $^{87}$Rb spin bath.  At zero transverse coherence, the $F=1$ spinor gas polarization is defined by the dimensionless longitudinal magnetization $M_z$ and quadrupole moment $Q_{zz}$, whose range is indicated by the purple shaded region.  The $^7$Li steady-state distribution is predicted to lie in a narrower region (shaded yellow) given all possible fixed polarizations of the $^{87}$Rb bath.  Measured late-time $^7$Li distributions are marked by orange markers with dashed boundary, while the predicted steady state distributions, based on measurements of the \Rb\ gas spin distributions, are shown with purple markers. The corresponding prepared \Rb\ gas spin distributions are close to the predicted $^7$Li steady state spin distributions\footnote{ For the three settings described in Fig. 3, the values $(M_z, Q_{zz})$ for the Rb bath are $(-0.44, 0)$ (diamond), $(0, 0)$ (star), and $(0.74, 0.19)$ (square and circle). The corresponding predictions for the long-time state of the Li gas, corresponding the locations of markers, are $(-0.42, 0.04)$, $(0, 0)$, and $(0.76, 0.17)$, respectively.}.  Four initial states are considered: (a) diamond, (b) circle, (c) star, and (Rb bath as in b, but Li initialized in $|m_F = +1\rangle$ state) square.  Here, the labels (a-c) refer to Fig.\ \ref{fig:LiRelaxInRbBath.pdf}.}
\label{fig:LiRbLongTimeEvo.pdf}
\end{center}
 \end{figure}

Such steady-state distributions, determined from the measured spin-dependent cross sections, are presented in Fig.\ \ref{fig:LiRbLongTimeEvo.pdf}.  Here, we characterize the spin distribution by its dimensionless longitudinal magnetization $M_z= p_{+1}- p_{-1}$ and one component of the dimensionless quadrupole tensor $Q_{zz} = \frac{1}{3} - p_0$. While the full range of spin distributions is described by $M_z$ ranging from -1 to +1 and $Q_{zz}$ ranging from $-\frac{2}{3}$ to $\frac{1}{3}$, the full range of steady-state Li distributions, calculated for all possible distributions of the Rb bath, is predicted to occupy a narrow band of possible values.  We test this prediction experimentally, considering the three Rb bath distributions shown in Fig.\ \ref{fig:LiRelaxInRbBath.pdf}, and comparing the observed long-time Li distribution to that predicted theoretically, finding good agreement between the two.

Over much longer times, the spin distribution of the Rb gas should also evolve \footnote{Given $N^{(\mathrm{Rb})}_\mathrm{tot}/N^{(\mathrm{Li})}_\mathrm{tot} \simeq 10$, the expected fractional change in the Rb gas magnetization (around 10\%) is below the resolution and stability limits of our measurement.}.  Considering just the effects of Li-Rb collisions, the timescale for the evolution of the spin-state distribution of the Rb gas is longer than that of the Li gas by the factor $N^{(\mathrm{Rb})}_\mathrm{tot}/N^{(\mathrm{Li})}_\mathrm{tot} \simeq 10$.  Over this long timescale, which was inaccessible in our experiment, the spin-state distributions of the two gases should acquire equal longitudinal magnetizations and quadrupole moments \cite{suppmat}.

In conclusion, by characterizing spin relaxation in $^7$Li-$^{87}$Rb gas mixtures, we obtain the first complete experimental determination of all spin-dependent interactions in a heteronuclear spinor gas. Our work highlights the richness of heteronuclear spinor gases.  Whereas for homonuclear spinor gases the longitudinal magnetization is constant (neglecting magnetic dipole-dipole interactions) and determined manually, for the heteronuclear system, each spinor gas may undergo significant variation by exchanging magnetization with the other. Thus, by using one spinor gas as a large magnetization reservoir, one may obtain a clearer picture of the thermodynamic behavior of the other spinor gas without artificial constraint.  This work also points to an   analogy between the dynamics of heteronuclear spinor gases and solid-state spintronics.  For example, future work could study the evolution of an itinerant spinor gas, e.g.\ Li with its light mass, in the presence of another spatially localized spinor gas, e.g.\ a heavier Rb or Cs gas, in order to simulate the evolution of electrons within layered magnetic materials.


We acknowledge support from the National Science Foundation (PHY-1707756), DTRA (HDTRA1-15-1-0009), and by NASA-JPL (JPL1505588). KH acknowledges support by 'Hamburglobal', funded by the program 'Promos' of the German Academic Exchange Service (DAAD), issued through Universität Hamburg. The work of CHG and YW has been supported by NSF grant PHY-1912350.

\bibliography{allrefs_x2.bib, references.bib, lirb_notes.bib, LiRbSuppNote.bib}

\begin{thebibliography}{25}
\bibitem [{ram()}]{ramseynote}%
  \BibitemOpen
  \href@noop {} {}\bibinfo {note} {A Ramsey-pulse scheme is used to observe the
  few-ms decay time of transverse magnetization.}\BibitemShut {Stop}%
\end{thebibliography}
\end{document}